\documentstyle[12pt]{article}

\input{tcilatex}

\begin{document}

\title{Angular Momentum of the BTZ Black Hole in the Teleparallel Geometry}
\author{A. A. Sousa*, R. B. Pereira \\
Departamento de Matem\'{a}tica\\
Instituto de Ci\^{e}ncias e Letras do M\'{e}dio Araguaia \\
Universidade Federal de Mato Grosso\\
78698-000 Pontal do Araguaia, MT, Brazil\\
Departamento de F\'{\i}sica, ICET-UFMT, MT, Brazil\\
and\\
J. F. da Rocha-Neto\\
Funda\c c\~ao Universidade Federal do Tocantins\\
Campus Universit\'ario de Arraias,\\
Rua Universit\'aria, s/n centro Arraias-TO, Brazil \\
77330-000\\
}
\maketitle

\begin{abstract}
We carry out the Hamiltonian formulation of the three-dimensional
gravitational teleparallelism without imposing the time gauge condition, by
rigorously performing the Legendre transform. Definition of the
gravitational angular momentum arises by suitably interpreting the integral
form of the constraint equation $\Gamma ^{ik}=0$ as an angular momentum
equation. The gravitational angular momentum is evaluated for the
gravitational field of a rotating BTZ black hole.

PACS NUMBERS: 04.20.Cv, 04.20.Fy

(*) E-mail: adellane@cpd.ufmt.br
\end{abstract}

\section{Introduction}

The search for a consistent expression for the gravitational energy and
angular momentum of a self-gravitating distribution of matter is undoubtedly
a long-standing problem in general relativity. The gravitational field does
not possess the proper definition of an energy momentum tensor and an
angular momentum tensor and one usually defines some energy-momentum and
angular momentum as Bergmann \cite{1} or Landau-Lifschitz \cite{2} which are
pseudo-tensors.The {\normalsize Einstein's general relativity can also be
reformulated in the context of the teleparallel (Weitzenb\"{o}ck) geometry.
In this geometrical setting the dynamical field quantities correspond to
orthornormal tetrad fields $e^{a}\,_{\mu }$ ($a,\,\mu $ are SO(3,1) and
space-time indices, respectively). These fields allow the construction of
the Lagrangian density of the teleparallel equivalent of general relativity
(TEGR), which offers an alternative geometrical framework for Einstein's
equations. The Lagrangian density for the tetrad field in the TEGR is given
by a sum of quadratic terms in the torsion tensor $T^{a}\,_{\mu \nu
}=\partial _{\mu }e^{a}\,_{\nu }-\partial _{\nu }e^{a}\,_{\mu }$, which is
related to the anti-symmetric part of Cartan's connection $\Gamma _{\mu \nu
}^{\lambda }=e^{a\lambda }\partial _{\mu }e_{a\nu }$. The curvature tensor
constructed out of the latter vanishes identically. This connection defines
a space with teleparallelism, or absolute parallelism \cite{3}.}

The {\normalsize Hamiltonian formulation, when consistently established, not
only guarantees that field quantities have a well defined time evolution,
but also allow us to understand physical theories from a different
perspective. We have learned from the work of Arnowitt, Deser and Misner
(ADM) \cite{4} that the Hamiltonian analysis of Einstein's general
relativity reveals the intrinsic structure of the theory: the time evolution
of field quantities is determined by the Hamiltonian and vector constraints.
Thus four of the ten Einstein's equations acquire a prominent status in the
Hamiltonian framework. Ultimately this is an essential feature for the
canonical approach to the quantum theory of gravity. In the framework of the
TEGR it is possible to make definite statements about the energy and
momentum of the gravitational field. This fact constitutes the major
motivation for considering this theory. In the 3+1 formulation of the TEGR 
\cite{5}, and by imposing Schwinger's time gauge condition \cite{6} for the
tetrad field by fixing }$e_{\left( k\right) }\hspace{0.25em}^{0}=0$ which
implies $e^{\left( 0\right) }\,_{k}=0${\normalsize , it is found that the
Hamiltonian and vector constraints contain each one a divergence in the form
of scalar and vector densities, respectively, that can be identified with
the energy and momentum {\it densities} of the gravitational field \cite{7}.
This identification has proved to be consistent, and it has been
demonstrated that teleparallel theories provide a natural instrument for the
investigation of gravitational energy. The most relevant application is the
evaluation of the irreducible mass of the Kerr black hole \cite{8}.}

{\normalsize In the Hamiltonian formulation of the TEGR, with no {\it a
priori} restriction on the tetrad fields, there arises a set of primary
constraints $\Gamma ^{ik}$ that satisfy the angular momentum algebra \cite{9}%
. Following the prescription for defining the gravitational energy, the
definition of the gravitational angular momentum arises by suitably
interpreting the integral form of the constraint equation $\Gamma ^{ik}=0$
as an angular momentum equation. It has been applied this definition to the
gravitational field of a thin, slowly rotating mass shell \cite{10}. }

On the other hand, gravity theories in three dimensions have gained
considerable attention in the recent years. The expectation is that the
study of lower-dimensional theories will provide relevant information about
the corresponding theory in four dimensions. A family of three-parameter
teleparallel theories in 2+1 dimensions are proposed by Kawai \cite{11}, and
a Hamiltonian formulation was developed \cite{12}. The definition of energy
of the gravitational field \cite{13} was proposed with the use of the 
{\normalsize orthornormal triads fields $e^{a}\,_{\mu }$ ($a,\,\mu $ are
SO(2,1) and space-time indices, respectively)} in Schwinger's time gauge 
\cite{6}.

{\normalsize In this paper we carry out the Hamiltonian formulation of this
three-parameter theory in three-dimensions without imposing the time gauge
condition, by rigorously performing the Legendre transform. We have not
found it necessary to establish a 2+1 decomposition for the triad field.}
Following Dirac's method \cite{14}, we find that the three-parameter family
is reduced to one-parameter family$.$ We conclude that the Legendre
transform is well defined only if certain conditions on the parameters are
satisfied. These conditions are $c_{1}+\frac{8}{3}c_{3}=0$ and $%
3c_{1}+4c_{2}=0$ (These results appeared in the canonical formulation by
imposing the time gauge condition)$.$ These are the conditions for the
formulation to be well defined from the point of view of the initial value
problem. The value of this unique parameter was fixed in previous works of
the author \cite{13} as -2/3, leading to the solution of the so-called
three-dimensional BTZ black hole solution \cite{15} in the teleparallel
geometry. Since its discovery in 1992, BTZ black hole solution has often
been used in current literature as a simple but realistic model for black
hole physics.

The BTZ solution has been used to study quantum models in 2+1 dimensions
because the corresponding models in 3+1 dimensions are very complicated \cite%
{16}. {\normalsize The constraint algebra of the theory suggests that
certain momentum components are related to the gravitational angular
momentum. It turns out to be possible to define, in this context, the
angular momentum of the gravitational field.} {\normalsize We apply this
definition to the gravitational field of a rotating BTZ black hole}. 
{\normalsize It turns out, however, that consistent values for the
gravitational angular momentum are achieved by requiring the triad field to
satisfy ({\it a posteriori}) the time gauge condition }(this seems to be a
requirement to obtain the irreducible mass of the Kerr black hole, in the
quadridimensional case) \cite{10}. In the region near to the surface of the
BTZ black hole $r>r_{0}$, we arrive at $M^{12}\simeq -\frac{J}{2}\left( 1-%
\frac{r_{0}}{r}\right) $ where $J$ is the angular momentum of the source and 
$r_{0}$ is the external horizon of the BTZ black hole. We compare our
expression of gravitational angular momentum with the result obtained by
means of Komar method {\normalsize (this approach assumes the existence of
certain Killing vector fields that allow the construction of conserved
integral quantities) }\cite{17}. We find that the results are different.
However, it is noted that the result obtained by means of our method is much
simpler and does not depend on Killing vectors. We confirm that we are
dealing with the angular momentum of the gravitational field and not of the
source of the field. These results indicate that the alternative geometrical
formulation provided by the TEGR allows to obtain expressions for the
energy, momentum and angular momentum of the gravitational field.

The article is divided as follows. In the section 2, we review the
Lagrangian formulation and carry out the Hamiltonian formulation of
arbitrary teleparallel theories without fixing gauge. In the section 3, 
{\normalsize we obtain a realistic measure of the angular momentum of the
field in terms of the angular momentum of the source}. In the section 4, we
compare our result with the gravitational angular momentum of BTZ black hole
calculated by means of the Komar method. Finally, in the section 5, we
present our conclusions.

We employ the following notation{\normalsize . Space-time indices $\mu ,\nu
,...$ and SO(2,1) indices $a,b,...$ run from 0 to 2. Time and space indices
are indicated according to $\mu =0,i,\hspace{0.75em}\hspace{0.75em}a=(0),(i)$%
. The triad field $e^{a}\hspace{0.25em}_{\mu }$ yields the definition of the
torsion tensor: $T^{a}\hspace{0.25em}_{\mu \nu }=\partial _{\mu }e^{a}%
\hspace{0.25em}_{\nu }-\partial _{\nu }e^{a}\hspace{0.25em}_{\mu }$. The
flat Minkowski space-time metric is fixed by $\eta _{ab}=e_{a\mu }e_{b\nu
}g^{\mu \nu }=(-++)$. }

\section{Hamiltonian formulation}

We begin by introducing the three basic postulates that the Lagrangian
density of the gravitational field in empty space, in the teleparallel
geometry, must satisfy. It must be invariant under (i) coordinate
transformations, (ii) global $\left[ SO(2,1)\right] $ Lorentz's
transformations, and (iii) parity transformations. In the present
formulation, we add a negative cosmological constant $\Lambda =-\frac{2}{%
l^{2}}$ to the Lagrangian density. The most general Lagrangian density
quadratic in the torsion tensor is written 
\begin{equation}
L=e\left( c_{1}t^{abc}t_{abc}+c_{2}v^{a}v_{a}+c_{3}a_{abc}a^{abc}\right)
\label{1}
\end{equation}%
where $c_{1}$, $c_{2}$ and $c_{3}$ are constants, $e=\det \left( e^{a}%
\hspace{0.25em}_{\mu }\right) ,$ and 
\begin{equation}
t_{abc}=\frac{1}{2}\left( T_{abc}+T_{bac}\right) +\frac{1}{4}\left( \eta
_{ca}v_{b}+\eta _{cb}v_{a}\right) -\frac{1}{2}\eta _{ab}v_{c}\text{ ,}
\end{equation}%
\begin{equation}
v_{a}=T_{\text{ \ }ba}^{b}\text{ ,}
\end{equation}%
\begin{equation}
a_{abc}=\frac{1}{3}\left( T_{abc}+T_{cab}+T_{bca}\right) \text{ ,}
\end{equation}%
\begin{equation}
T_{abc}=e_{b}^{\text{ \ }\mu }e_{c}^{\text{ \ }\nu }T_{a\mu \nu }\text{ .}
\end{equation}%
The definitions given above correspond to the irreducible components of the
torsion tensor \cite{11}. The field equations for the Lagrangian density $%
\left( \ref{1}\right) $ are obtained in Ref. \cite{11}.

The Hamiltonian formulation is obtained by writing the Lagrangian density in
first-order differential form.{\normalsize \ For this purpose we introduce
an auxiliary field quantity $\Delta _{abc}=-\Delta _{acb}$ that will be
related to the torsion tensor. The first-order differential Lagrangian
formulation in empty space-time reads } 
\begin{equation}
L(e,\Delta )\hspace{0.75em}=\hspace{0.75em}-e\Lambda ^{abc}\left( \Delta
_{abc}-2T_{abc}\right) \text{{}},  \label{2}
\end{equation}%
where 
\[
\Lambda ^{abc}=c_{1}\Theta ^{abc}+c_{2}\Omega ^{abc}+c_{3}\Gamma ^{abc}, 
\]%
and $\Theta ^{abc},$ $\Omega ^{abc}$ and $\Gamma ^{abc}$ are defined as, 
\begin{equation}
\Theta ^{abc}=\frac{1}{2}\Delta ^{abc}+\frac{1}{4}\Delta ^{bac}-\frac{1}{4}%
\Delta ^{cab}+\frac{3}{16}\left( \eta ^{ca}\Delta ^{b}-\eta ^{ba}\Delta
^{c}\right) \text{ ,}
\end{equation}%
\begin{equation}
\Omega ^{abc}=\frac{1}{2}\left( \eta ^{ab}\Delta ^{c}-\eta ^{ac}\Delta
^{b}\right) \text{ ,}
\end{equation}%
\begin{equation}
\Gamma ^{abc}=\frac{1}{3}\left( \Delta ^{abc}+\Delta ^{bca}+\Delta
^{cab}\right) \text{ ,}
\end{equation}%
{\normalsize where $T_{abc}=e_{b}\hspace{0.25em}^{\mu }e_{c}\hspace{0.25em}%
^{\nu }T_{a\mu \nu }$ and $\Delta _{b}=\Delta ^{a}\hspace{0.25em}_{ab}$.}

{\normalsize Variation of the action constructed out of (\ref{2}) with
respect to $\Delta ^{abc}$ yields an equation that can be reduced to $\Delta
_{abc}\hspace{0.75em}=\hspace{0.75em}T_{abc}$. This equation can be split
into two equations: }

{\normalsize 
\[
\Delta_{a 0 k} \hspace{0.75em} = \hspace{0.75em} T_{a 0 k} \hspace{0.75em} = 
\hspace{0.75em} \partial_0 e_{ak} - \partial_k e_{a 0} \hspace{0.75em}, 
\]%
}

{\normalsize 
\begin{equation}
\Delta _{aik}\hspace{0.75em}=\hspace{0.75em}T_{aik}\hspace{0.75em}=\hspace{%
0.75em}\partial _{i}e_{ak}-\partial _{k}e_{ai}\hspace{0.75em}.  \label{8000}
\end{equation}%
}

{\normalsize The Hamiltonian density will be obtained by the standard
prescription $L=p\dot{q}-H_{0}$ and by properly identifying primary
constraints. We have not found it necessary to establish any kind of 2+1
decomposition for the triad field. Therefore in the following both $e_{a\mu
} $ and $g_{\mu \nu }$ are space-time fields. The analysis developed here is
similar to that developed in Refs. \cite{9,10}. }

The {\normalsize Lagrangian density can be expressed as } 
\begin{equation}
L(e,\phi )\hspace{0.75em}=\hspace{0.75em}4e\hspace{0.25em}\Lambda ^{a0k}%
\hspace{0.25em}\dot{e}_{ak}-4e\hspace{0.25em}\Lambda ^{a0k}\hspace{0.25em}%
\partial _{k}e_{a0}+2e\hspace{0.25em}\Lambda ^{aij}\hspace{0.25em}%
T_{aij}+e\Lambda ^{abc}\hspace{0.25em}\phi _{abc}\hspace{0.75em},
\label{5000}
\end{equation}%
where the dot indicates time derivative, and $\Lambda ^{a0k}=\Lambda ^{abc}%
\hspace{0.25em}e_{b}\hspace{0.25em}^{0}\hspace{0.25em}e_{c}\hspace{0.25em}%
^{k}$, $\Lambda ^{aij}=\Lambda ^{abc}\hspace{0.25em}e_{b}\hspace{0.25em}^{i}%
\hspace{0.25em}e_{c}\hspace{0.25em}^{j}$.

{\normalsize Therefore the momentum canonically conjugated to $e_{ak}$ is
given by } 
\begin{equation}
P^{ak}\hspace{0.75em}=\hspace{0.75em}4\hspace{0.25em}e\hspace{0.25em}\Lambda
^{a0k}\hspace{0.75em}.  \label{6000}
\end{equation}%
In terms of (\ref{6000}) expression (\ref{5000}) reads 
\begin{equation}
L=P^{ak}\hspace{0.25em}\dot{e}_{ak}-P^{ak}\hspace{0.25em}\partial
_{k}e_{a0}-e\Lambda ^{aij}(-2T_{aij}+\Delta _{aij})-2e\Lambda ^{a0i}\Delta
_{a0i}\hspace{0.75em}.  \label{7000}
\end{equation}%
The last term on the right hand side of equation (\ref{7000}) is identified
as

\[
2e\Lambda ^{a0i}\Delta _{a0i}=\frac{1}{2}P^{ai}\Delta _{a0i}. 
\]

{\normalsize The Hamiltonian formulation is established once we rewrite the
Lagrangian density (\ref{7000}) in terms of $e_{ak}$, $\Pi ^{ak}$ and
further nondynamical field quantities. It is carried out in two steps.
First, we take into account equations (\ref{8000}) and (\ref{7000}) so that
half of the auxiliary fields, $\phi _{aij}$, are eliminated from the
Lagrangian density by means of the identification} 
\begin{equation}
\Delta _{aij}=T_{aij}\hspace{0.75em}.
\end{equation}%
Therefore we have

\begin{equation}
L(e_{ak},P^{ak},e_{a0},\Delta _{a0k})\hspace{0.75em}=\hspace{0.75em}P^{ak}%
\dot{e}_{ak}+e_{a0}\partial _{k}P^{ak}-\partial _{k}(e_{a0}P^{ak})
\end{equation}

\[
-e\biggl[\frac{1}{4}g^{ik}g^{jl}T^{a}\hspace{0.25em}_{kl}T_{aij}\left( \frac{%
c_{1}}{2}+\frac{c_{3}}{3}\right) +\left( -\frac{1}{2}c_{1}+\frac{2}{3}%
c_{3}\right) g^{il}T^{j}\hspace{0.25em}_{kl}T^{k}\hspace{0.25em}_{ij}+ 
\]

\[
+\left( -\frac{3}{4}c_{1}+c_{2}\right) g^{ik}T^{j}\hspace{0.25em}_{ji}T^{n}%
\hspace{0.25em}_{nk}\biggr ]
\]%
\[
-\frac{1}{2}\Delta _{a0k}\biggl\{P^{ak}-2e\hspace{0.25em}\biggl[%
g^{i0}g^{jk}T^{a}\hspace{0.25em}_{ij}\left( c_{1}+\frac{2}{3}c_{3}\right) + 
\]%
\[
+e^{ai}(g^{j0}T^{k}\hspace{0.25em}_{ij}-g^{jk}T^{0}\hspace{0.25em}%
_{ij})\left( -\frac{1}{2}c_{1}+\frac{2}{3}c_{3}\right) + 
\]%
\[
\left( \frac{3}{4}c_{1}-c_{2}\right) (e^{ak}g^{i0}-e^{a0}g^{ik})T^{j}\hspace{%
0.25em}_{ji}\biggr ]\biggr\}. 
\]

\subsection{The canonical momentum}

\QTP{ite}
{\normalsize The second step consists of expressing the remaining auxiliary
field quantities, the \textquotedblleft velocities\textquotedblright\ $%
\Delta _{a0k}$, in terms of the momenta $P^{ak}$. This is the nontrivial
step of the Legendre transform. Denoting $(..)$ and $[..]$ as the symmetric
and anti-symmetric parts of field quantities, respectively, we decompose $%
P^{ak}$ into irreducible components plus terms of excess: }

\QTP{ite}
{} {\normalsize 
\begin{eqnarray}
P^{ak}\hspace{0.75em} &=&\hspace{0.75em}e^{a}\hspace{0.25em}_{i}\hspace{%
0.25em}P^{(ik)}+e^{a}\hspace{0.25em}_{i}\hspace{0.25em}P^{[ik]}+e^{a}\hspace{%
0.25em}_{0}\hspace{0.25em}P^{0k}+e\biggl\{e^{a}\hspace{0.25em}_{i}\biggl\{%
-\left( c_{1}+\frac{8}{3}c_{3}\right) g^{00}g^{ik}\Delta ^{j}\hspace{0.25em}%
_{0j}  \nonumber \\
&&+\frac{1}{2}\left( 3c_{1}+4c_{2}\right) g^{00}g^{ik}\Delta ^{j}\hspace{%
0.25em}_{0i}+  \nonumber \\
&&+\left( c_{1}+\frac{8}{3}c_{3}\right) g^{0i}g^{0k}\Delta ^{j}\hspace{0.25em%
}_{0j}-\frac{1}{2}\left( 3c_{1}+4c_{2}\right) g^{0i}g^{0k}\Delta ^{j}\hspace{%
0.25em}_{0j}+  \nonumber \\
&&+\left( c_{1}+\frac{8}{3}c_{3}\right) g^{ik}g^{0j}\Delta ^{0}\hspace{0.25em%
}_{0j}-\frac{1}{2}\left( 3c_{1}+4c_{2}\right) g^{ik}g^{0j}\Delta ^{0}\hspace{%
0.25em}_{0j}+  \nonumber \\
&&+\left[ \left( c_{1}+\frac{8}{3}c_{3}\right) -\frac{1}{2}\left(
3c_{1}+4c_{2}\right) \right] g^{00}g^{ik}T^{j}\hspace{0.25em}_{jl}\biggr\}+ 
\nonumber \\
&&+\left( c_{1}+\frac{8}{3}c_{3}\right) g^{00}g^{jk}\Delta ^{i}\hspace{0.25em%
}_{0j}-\left( c_{1}+\frac{8}{3}c_{3}\right) g^{0k}g^{j0}T^{i}\hspace{0.25em}%
_{0j}+  \nonumber \\
&&+\frac{1}{2}\left[ \left( 3c_{1}+4c_{2}\right) -\left( c_{1}+\frac{8}{3}%
c_{3}\right) \right] g^{0i}g^{kj}T^{0}\hspace{0.25em}_{0j}+  \nonumber \\
&&+\left( c_{1}+\frac{8}{3}c_{3}\right) g^{l0}g^{jk}T^{i}\hspace{0.25em}%
_{lj}-  \nonumber \\
&&-\left[ \left( c_{1}+\frac{8}{3}c_{3}\right) -\frac{1}{2}\left(
3c_{1}+4c_{2}\right) \right] g^{0i}g^{kl}T^{j}\hspace{0.25em}_{jl}  \nonumber
\\
&&+e^{a}\hspace{0.25em}_{0}\biggl\{\frac{1}{2}\left[ \left(
3c_{1}+4c_{2}\right) -\left( c_{1}+\frac{8}{3}c_{3}\right) \right]
g^{00}g^{ki}\Delta ^{0}\hspace{0.25em}_{0i}-  \nonumber \\
&&-\frac{1}{2}\left( 3c_{1}+4c_{2}\right) g^{0k}g^{0i}\Delta ^{0}\hspace{%
0.25em}_{0i}+\left( c_{1}+\frac{8}{3}c_{3}\right)
(g^{00}g^{kl}-g^{0l}g^{0k})T^{j}\hspace{0.25em}_{lj}  \nonumber \\
&&+\frac{1}{2}\left( 3c_{1}+4c_{2}\right) (g^{0l}g^{0k}-g^{00}g^{kl})T^{j}%
\hspace{0.25em}_{lj}+  \nonumber \\
&&\left( c_{1}+\frac{8}{3}c_{3}\right) g^{00}g^{ik}\Delta ^{0}\hspace{0.25em}%
_{0i}+\left( c_{1}+\frac{8}{3}c_{3}\right) g^{i0}g^{jk}T^{0}\hspace{0.25em}%
_{ij}\biggr\}\biggr\},  \label{2001}
\end{eqnarray}%
where }

\QTP{}
{\normalsize 
\begin{eqnarray*}
P^{(ik)}\hspace{0.75em} &=&\hspace{0.75em}-\left( c_{1}-\frac{4}{3}%
c_{3}\right) \hspace{0.25em}e\biggl\{g^{00}(-g^{jk}\Delta ^{i}\hspace{0.25em}%
_{0j}-g^{ji}\Delta ^{k}\hspace{0.25em}_{0j}+2g^{ik}\Delta ^{j}\hspace{0.25em}%
_{0j})+ \\
&&+g^{0k}(g^{j0}\Delta ^{i}\hspace{0.25em}_{0j}+g^{ji}\Delta ^{0}\hspace{%
0.25em}_{0j}-g^{0i}\Delta ^{j}\hspace{0.25em}_{0j})
\end{eqnarray*}%
}

\QTP{}
{\normalsize 
\begin{equation}
+g^{0i}(g^{0j}\Delta ^{k}\hspace{0.25em}_{0j}+g^{kj}\Delta ^{0}\hspace{0.25em%
}_{0j}-g^{0k}\Delta ^{j}\hspace{0.25em}_{0j})-2g^{ik}\hspace{0.25em}%
g^{0j}\Delta ^{0}\hspace{0.25em}_{0j}\hspace{0.75em}+\hspace{0.75em}N^{ik}%
\biggr\}\hspace{0.75em},
\end{equation}%
}

\QTP{}
{\normalsize 
\begin{equation}
N^{ik}\hspace{0.75em}=\hspace{0.75em}-g^{0l}(g^{jk}T^{i}\hspace{0.25em}%
_{lj}+g^{ji}T^{k}\hspace{0.25em}_{lj}-2g^{ik}T^{j}\hspace{0.25em}%
_{lj})-(g^{kl}g^{0i}+g^{il}g^{0k})T^{j}\hspace{0.25em}_{lj}\hspace{0.75em},
\end{equation}%
}

\QTP{}
{\normalsize 
\begin{equation}
P^{[ik]}\hspace{0.75em}=\hspace{0.75em}-\hspace{0.25em}\left( c_{1}-\frac{4}{%
3}c_{3}\right) e\left\{ -g^{li}g^{kj}T^{0}\hspace{0.25em}%
_{lj}+(g^{il}g^{0k}-g^{kl}g^{0i})T^{j}\hspace{0.25em}_{lj}\right\} \hspace{%
0.75em},  \label{9000}
\end{equation}%
} 
\begin{equation}
P^{0k}\hspace{0.75em}=\hspace{0.75em}2\left( c_{1}-\frac{4}{3}c_{3}\right) 
\hspace{0.25em}e\hspace{0.25em}\left[ g^{i0}g^{jk}T^{0}\hspace{0.25em}%
_{ij}+(g^{00}g^{kl}-g^{0l}g^{0k})T^{j}\hspace{0.25em}_{lj})\right] \hspace{%
0.75em}.  \label{10000}
\end{equation}%
where we have already identified $\Delta _{aij}=T_{aij}$.

\subsection{Conditions on the free parameters}

Before carrying out the Legendre transform we can establish the conditions
under which the Lagrangian density will be exempt of {\normalsize the
\textquotedblleft velocities\textquotedblright\ $\Delta _{a0k}.$ }We see
that the excess terms, which contains several {\normalsize $\Delta _{a0k}$
type terms, is discarded if we require }%
\begin{eqnarray}
c_{1}+\frac{8}{3}c_{3} &=&0,  \label{2000} \\
3c_{1}+4c_{2} &=&0.  \nonumber
\end{eqnarray}%
These results also appeared in the canonical formulation imposing the time
gauge condition \cite{12}. {\normalsize The crucial point in this analysis
is that only the symmetrical components $P^{(ij)}$ depend on the
\textquotedblleft velocities\textquotedblright\ $\Delta _{a0k}$. The other
three components, $P^{[ij]}$ and $P^{0k}$ depend solely on $T_{aij}$.
Therefore we can express only three of the \textquotedblleft
velocity\textquotedblright\ fields $\Delta _{a0k}$ in terms of the
components $P^{(ij)}$. With the purpose of finding out which components of $%
\Delta _{a0k}$ can be inverted in terms of the momenta we decompose $\Delta
_{a0k}$ identically as } 
\begin{equation}
\Delta ^{a}\hspace{0.25em}_{0j}\hspace{0.75em}=\hspace{0.75em}e^{ai}\hspace{%
0.25em}\psi _{ij}+e^{ai}\hspace{0.25em}\sigma _{ij}+e^{a0}\hspace{0.25em}%
\lambda _{j}\hspace{0.75em},  \label{3000}
\end{equation}%
where $\psi _{ij}=\frac{1}{2}(\Delta _{i0j}+\Delta _{j0i})$,$\hspace{0.75em}%
\hspace{0.75em}$ $\sigma _{ij}=\frac{1}{2}(\Delta _{i0j}-\Delta _{j0i})$,$%
\hspace{0.75em}\hspace{0.75em}$ $\lambda _{j}=\Delta _{00j}$,$\hspace{0.75em}%
\hspace{0.75em}$ and $\Delta _{\mu 0j}=e^{a}\hspace{0.25em}_{\mu }\Delta
_{a0j}$ (like $\Delta _{abc}$, the components $\psi _{ij}$, $\sigma _{ij}$
and $\lambda _{j}$ are also auxiliary field quantities). By defining 
\begin{equation}
\Pi ^{ik}\hspace{0.75em}=\hspace{0.75em}\frac{1}{e}P^{(ik)}+\left( c_{1}-%
\frac{4}{3}c_{3}\right) M^{ik}\hspace{0.75em},
\end{equation}%
we find that $\Pi ^{ik}$ depends only on $\psi _{ij}$:

\QTP{}
{\normalsize 
\[
\Pi ^{ik}\hspace{0.75em}=\hspace{0.75em}-2g^{00}(g^{il}g^{jk}\psi
_{lj}-g^{ik}\psi )+ 
\]%
} 
\begin{equation}
+2(g^{0i}g^{kl}g^{0j}+g^{0k}g^{il}g^{j0})\psi _{lj}-2(g^{ik}g^{0l}g^{0j}\psi
_{lj}+g^{0i}g^{0k}\psi )\hspace{0.75em},
\end{equation}%
where $\psi =g^{ik}\psi _{ik}$.

\QTP{}
{\normalsize We can now invert $\psi _{mn}$ in terms of $\Pi ^{ik}$. After a
number of manipulations we arrive at } 
\begin{equation}
\psi _{mn}\hspace{0.75em}=\hspace{0.75em}-\frac{1}{2g^{00}}\left(
g_{im}g_{kn}\Pi ^{ik}-g_{mn}\hspace{0.25em}\Pi \right) \hspace{0.75em},
\label{4000}
\end{equation}%
where $\Pi =g_{ik}\Pi ^{ik}$.

{\normalsize At last we need to rewrite the third line of the Lagrangian
density in terms of canonical variables. By making use of (\ref{2001}), (\ref%
{2000}), (\ref{3000}) and }$\left( {\normalsize \ref{4000}}\right) $ 
{\normalsize we can rewrite }

{\normalsize 
\[
-\frac{1}{2}\Delta _{a0k}\biggl\{P^{ak}-\frac{3}{2}ec_{1}\biggl\{%
g^{0i}g^{jk}T^{a}\hspace{0.25em}_{ij}- 
\]%
} 
\[
-e^{ai}(g^{0j}T^{k}\hspace{0.25em}_{ij}-g^{kj}T^{0}\hspace{0.25em}_{ij})+ 
\]%
\[
+2(e^{ak}g^{0i}-e^{a0}g^{ki})T^{j}\hspace{0.25em}_{ji}\biggr\}\biggr\}= 
\]%
\[
-\frac{1}{2}\Delta _{a0k}\biggl\{-\frac{3}{2}ec_{1}\biggl\{%
g^{00}(-g^{ik}\Delta ^{a}\hspace{0.25em}_{0i}-e^{ai}\Delta ^{k}\hspace{0.25em%
}_{0i}+2e^{ak}\Delta ^{i}\hspace{0.25em}_{0i})+ 
\]%
\begin{eqnarray}
&&+g^{0k}(g^{i0}\Delta ^{a}\hspace{0.25em}_{0i}+e^{ai}\Delta ^{0}\hspace{%
0.25em}_{0i})\hspace{0.25em}+e^{a0}(g^{0i}\Delta ^{k}\hspace{0.25em}%
_{0i}+g^{ki}\Delta ^{0}\hspace{0.25em}_{0i})- \\
&&-2(e^{a0}g^{k0}\Delta ^{i}\hspace{0.25em}_{0i}+e^{ak}g^{0i}\Delta ^{0}%
\hspace{0.25em}_{0i})\biggr\}\biggr\}\hspace{0.75em},  \nonumber
\end{eqnarray}%
in the form

\QTP{}
{\normalsize 
\begin{equation}
-\frac{3}{2}ec_{1}\left( \frac{1}{4g^{00}}\right) \left( g_{mi}g_{nj}\Pi
^{mn}\Pi ^{ij}-\Pi ^{2}\right) \hspace{0.75em}.
\end{equation}%
}

\subsection{Total Hamiltonian density}

\QTP{}
{\normalsize Thus we finally obtain the primary Hamiltonian density $%
H_{0}=P^{ak}\dot{e}_{ak}-L$, }

\QTP{}
{\normalsize 
\[
H_0 ( e_{ak}, P^{ak}, e_{a 0} ) \hspace{0.75em} = \hspace{0.75em} e_{a 0}
\partial_k P^{ak} + \frac{3}{2} ec_1 \left( \frac{1}{4 g^{00}} \right)
\left( g_{mi} g_{nj} \Pi^{mn} \Pi^{ij} - \Pi^2 \right) - 
\]%
}

\QTP{}
{\normalsize 
\begin{equation}
- \frac{3}{2} ec_1 \left( \frac{1}{4} g^{ik} g^{jl} T^a \hspace{0.25em}_{kl}
T_{aij} - \frac{1}{2} g^{il} T^j \hspace{0.25em}_{kl} T^k \hspace{0.25em}%
_{ij} - g^{ik} T^j \hspace{0.25em}_{ji} T^n \hspace{0.25em}_{nk} \right) 
\hspace{0.75em} .
\end{equation}%
}

\QTP{}
{\normalsize We may now write the total Hamiltonian density. For this
purpose we have to identify the primary constraints. They are given by
expressions (\ref{9000}) and (\ref{10000}), which represent relations
between $e_{ak}$ and the momenta $\Pi ^{ak}$. Thus we define (primary
constraints)}

{\normalsize 
\begin{equation}
\Gamma ^{ik}\hspace{0.75em}=\hspace{0.75em}-\Gamma ^{ki}\hspace{0.75em}=%
\hspace{0.75em}P^{[ik]}\hspace{0.75em}+\frac{3}{2}ec_{1}\hspace{0.25em}%
e\left\{ -g^{il}g^{kj}T^{0}\hspace{0.25em}%
_{lj}+(g^{il}g^{0k}-g^{kl}g^{0i})T^{j}\hspace{0.25em}_{lj}\right\} \hspace{%
0.75em},
\end{equation}%
} 
\begin{equation}
{\normalsize \Gamma ^{k}\hspace{0.75em}=\hspace{0.75em}P^{0k}\hspace{0.75em}%
-3ec_{1}\hspace{0.25em}e\hspace{0.25em}\left[ g^{jk}g^{0i}T^{0}\hspace{0.25em%
}_{ij}+\left( g^{00}g^{kl}-g^{0l}g^{0k}T^{j}\hspace{0.25em}_{ij}\right) T^{j}%
\hspace{0.25em}_{lj}\right] .}
\end{equation}%
{\normalsize Therefore the total Hamiltonian density is given by} {\bigskip }
\begin{equation}
H(e_{ak},P^{ak},e_{a0},\alpha _{ik},\beta _{k})=H_{0}+\alpha _{ik}\Gamma
^{ik}+\beta _{k}\Gamma ^{k}+\partial _{k}(e_{a0}P^{ak})\hspace{0.75em},
\end{equation}%
{\normalsize where $\alpha _{ik}$ and $\beta _{k}$ are Lagrange multipliers.}

\section{Algebra of the gravitational angular momentum}

\QTP{}
{\normalsize The calculations of the Poisson brackets between these
constraints are exceedingly complicated. We describe here only the algebra
of angular momentum. The Poisson bracket algebra closes in quadridimensional
formulation, since all the constraints of the theory are first class. }

\QTP{}
{\normalsize The Poisson bracket between two quantities $F$ and $G$ is
defined by } 
\[
\{F,G\}=\int d^{3}x\left( \frac{\delta F}{\delta e_{ai}(x)}\frac{\delta G}{%
\delta P^{ai}(x)}-\frac{\delta F}{\delta P^{ai}(x)}\frac{\delta G}{\delta
e_{ai}(x)}\right) \hspace{0.75em}, 
\]%
by means of which we can write down the evolution equations.

\QTP{}
{\normalsize The Hamiltonian density determines the time evolution of any
field quantity $f ( x )$: }

\QTP{}
{\normalsize 
\begin{equation}
\dot{f}(x)=\int d^{3}y\{f(x),H(y)\}|_{\Gamma ^{ik}=\Gamma ^{k}=0}\hspace{%
0.75em}.
\end{equation}%
}

\QTP{}
The Poisson brackets of primary constraints $\Gamma ^{ik}$ are given by 
\begin{equation}
\{\Gamma ^{ij}(x),\Gamma ^{kl}(y)\}={\frac{1}{2}}\left( g^{il}\Gamma
^{jk}+g^{jk}\Gamma ^{il}-g^{ik}\Gamma ^{jl}-g^{jl}\Gamma ^{ik}\right) \delta
(x-y)\hspace{0.75em},
\end{equation}%
that resemble the ordinary algebra of angular momentum.

\QTP{}
{\normalsize Finally we would like to remark that the Hamiltonian
formulation of the theory can be described more succinctly in terms of the
constraints $H_{0}$, $\Gamma ^{ik}$ and $\Gamma ^{k}$, by the Hamiltonian
density in the form }

\QTP{}
{\normalsize 
\begin{equation}
H(e_{ak},P^{ak},e_{a0},\alpha _{ik},\beta _{k})=H_{0}+\alpha _{ik}\Gamma
^{ik}+\beta _{k}\Gamma ^{k}\hspace{0.75em}.
\end{equation}%
}without the surface term.

{\bigskip}

\section{Gravitational angular momentum of a rotating BTZ black hole}

\QTP{}
{\normalsize The main motivation for considering the angular momentum of the
gravitational field in the present investigation resides in the fact that
the constraints $\Gamma ^{ik}$, satisfy the algebra of angular momentum.
Indeed, Ref. \cite{10} presented some progress in this respect.}

\QTP{}
{\normalsize Following the prescription for defining the gravitational
energy out of the Hamiltonian constraint of the TEGR and considering } $%
{\normalsize c}_{1}=-2/3,$ {\normalsize we interpret the integral form of
the constraint equation $\Gamma ^{ik}=0$ as an angular momentum equation,
and therefore we define the angular momentum of the gravitational field $%
M^{ik}$ according to } 
\begin{equation}
M^{ik}=\int_{S}d^{2}x\hspace{0.25em}P^{[ik]}=\int_{S}d^{2}x\hspace{0.25em}e%
\left[ -g^{li}g^{kj}T^{0}\hspace{0.25em}%
_{lj}+(g^{il}g^{0k}-g^{kl}g^{0i})T^{j}\hspace{0.25em}_{lj}\right] \hspace{%
0.75em},
\end{equation}%
for an arbitrary surface $S$ {\normalsize of the bi-dimensional space. }

\subsection{The determination of triad fields of the BTZ black hole}

\QTP{}
{\normalsize Since the definition of the above equation is a bi-dimensional
integral we will consider a space-time metric that exhibits rotational
motion. One exact solution that is everywhere regular in the exterior region
of the rotating source is the metric associated to a rotational BTZ black
hole. The main motivation for considering this metric is the construction of
a source for the exterior region of space-time. The metric reads \cite{15}} 
\begin{equation}
ds^{2}=-N^{2}dt^{2}+f^{-2}dr^{2}+r^{2}(N^{\theta }dt+d\theta )^{2},
\end{equation}%
where

\QTP{}
{\normalsize 
\begin{equation}
N(r)=f(r)=\sqrt{\left( -8GM+\frac{r^{2}}{l^{2}}+\frac{16G^{2}J^{2}}{r^{2}}%
\right) }\hspace{0.75em},
\end{equation}%
} 
\begin{equation}
N^{\theta }(r)=-\frac{4G}{r^{2}}J,
\end{equation}%
$J$ can be identified with the angular momentum of the source. T{\normalsize %
he gravitational constant }$G${\normalsize \ has the dimensions of an
inverse mass \cite{15}. The space-time geometry of the BTZ\ black hole is
one of constant negative curvature and therefore it is, locally, that of
anti-de Sitter (Ads) space. Thus, the BTZ black hole can only differ from
the anti-de Sitter space in its global properties.}

\QTP{}
{\normalsize The set of triad fields that satisfy the metric is given by } 
\begin{equation}
e_{a\mu }=\left( 
\begin{array}{ccc}
\left( -8GM+\frac{r^{2}}{l^{2}}+\frac{16G^{2}J^{2}}{r^{2}}\right) ^{1/2} & 0
& 0 \\ 
0 & \left( -8GM+\frac{r^{2}}{l^{2}}+\frac{16G^{2}J^{2}}{r^{2}}\right) ^{-1/2}
& 0 \\ 
-\frac{4GJ}{r} & 0 & r%
\end{array}%
\right) ,
\end{equation}%
where the determinant of $e_{a\mu }$ is 
\begin{equation}
e{\normalsize =r.}
\end{equation}

\QTP{}
T{\normalsize he only nonvanishing component of the torsion tensor that is
needed in the following read }

\QTP{}
{\normalsize 
\begin{equation}
T^{(2)}\hspace{0.25em}_{12}=1\hspace{0.75em}.
\end{equation}%
}

\QTP{}
{\normalsize The anti-symmetric components $P^{[ik]}$ can be easily
evaluated. We obtain }

\QTP{}
{\normalsize 
\begin{equation}
P^{[12]}(r)=eT^{(2)}\hspace{0.25em}_{12}g^{02}e_{\left( 2\right) }^{\text{ \
\ \ }2}g^{11},
\end{equation}%
}where 
\begin{eqnarray*}
g^{02} &=&-\frac{4GJ}{r^{2}\left( -M+\frac{r^{2}}{l^{2}}+\frac{16G^{2}J^{2}}{%
r^{2}}\right) }, \\
e_{\left( 2\right) }^{\text{ \ \ \ }2} &=&\frac{1}{r}, \\
g^{11} &=&\left( -8GM+\frac{r^{2}}{l^{2}}+\frac{16G^{2}J^{2}}{r^{2}}\right) .
\end{eqnarray*}%
{\normalsize The only nonvanishing component of the angular momentum is
given by}

\QTP{}
{\normalsize 
\begin{equation}
M^{12}=\frac{1}{16\pi G}\int_{S}d^{2}x\hspace{0.25em}P^{[12]}=-\frac{1}{4\pi 
}\int_{0}^{2\pi }d\theta \int_{r_{0}}^{r}\frac{J}{r}dr.
\end{equation}%
}The integral is calculated in the region near to the surface of the BTZ$\ $%
black hole $r>r_{0}$ \cite{18}$,$ because the metric possess {\normalsize %
regular solution in the exterior region of the rotating source. In the limit 
}$r\rightarrow r_{0},$ the {\normalsize calculation is straightforward.
Since }$\left\vert \frac{r_{0}}{r}-1\right\vert <1,$ {\normalsize we find }%
\begin{equation}
M^{12}=\frac{J}{2}\ln \left( 1+\frac{r_{0}-r}{r}\right) \simeq -\frac{J}{2}%
\left( 1-\frac{r_{0}}{r}\right) \hspace{0.75em},
\end{equation}%
where $r_{0}$ is the external horizon of the BTZ black hole. {\normalsize We
identify $M^{12}$ as the angular momentum of the gravitational field. }This
quantity is less\ than the angular momentum of the source. In the limit $%
r\rightarrow \infty $, we recover the anti-de Sitter space and $%
M^{12}\rightarrow -\infty $, because the gravitational field is more intense
at points far from the black hole. Moreover we know that the energy density
of the gravitational field diverges in the limit $r\rightarrow \infty $ \cite%
{19}.

\section{Integral of Komar}

\QTP{}
{\normalsize In order to assess the significance of the above result, we
present here the angular momentum associated to the metric tensor by means
of Komar's integral $Q_{K}$ \cite{17}, } 
\begin{equation}
Q_{K}=\frac{1}{8\pi }\oint_{S}\sqrt{-g}\hspace{0.25em}\varepsilon _{\alpha
\beta \mu \nu }\nabla ^{\lbrack \alpha }\xi ^{\beta ]}dx^{\mu }\wedge
dx^{\nu }\hspace{0.75em},
\end{equation}%
where $S$ {\normalsize is a surface of the \textquotedblleft
disk\textquotedblright\ of radius $R\rightarrow \infty $, $\xi ^{\mu }$ is
the Killing vector field $\xi ^{\mu }=\delta _{2}^{\mu }$ and $\nabla $ is
the covariant derivative constructed out of the Christoffel symbols $\Gamma
_{\mu \nu }^{\lambda }$. Introducing the gravitational constant }$G$, 
{\normalsize the integral $Q_{K}$ reduces to} 
\begin{equation}
Q_{K}=-\frac{J}{2}.
\end{equation}%
where $J$ is again the angular momentum of the source.

\QTP{}
{\normalsize One expects the gravitational angular momentum to be of the
order of magnitude of the intensity of the gravitational field. We observe
that Komar's integral yields a value proportional to the angular momentum of
the {\it source}, whereas $M^{12}$ is smaller than $Q_{K}$. In similarity to
Ref. \cite{10}, we observe that $M^{12}$ yields the angular momentum of the
gravitational field, not of the source, in contrast to $Q_{K}.$}

\section{Conclusions}

{\normalsize In the context of Einstein's general relativity rotational
phenomena is certainly not a completely understood issue. The prominent
manifestation of a purely relativistic rotational effect is the dragging of
inertial frames. If the angular momentum of the gravitational field of
isolated systems has as meaningful notion, than it is reasonable to expect
the latter to be somehow related to the rotational motion of the sources,
but not equal to the angular momentum of the sources. }The Hamiltonian
formulation of TEGR in three dimensions can provide an easy expression to
calculate the gravitational angular momentum with the use of the triad
fields obtained from the metric. It resulted to be smaller than Komar's
expression in the region near the black hole. Komar's expression is
independent of the surface of integration, and therefore its value does
represent the angular momentum of the source, not of the field{\normalsize . 
}It is necessary to test our expression of angular momentum to other
configurations of the gravitational field, to verify its consistency.
Efforts in this respect will be carried out.

\QTP{}
\bigskip \noindent {\it Acknowledgements}

\QTP{}
\noindent One of us (A. A. S.) would like to thank J. W. Maluf for the
carefully reading of the manuscript and comments. This work was partially
supported by the Brazilian agencies{\normalsize \ }CNPq and FAPEMAT. Grants
\#478997/03-5 and \#549/04, respectively.

\end{document}